\documentclass[12pt,english]{article}
\usepackage[T1]{fontenc}
\usepackage[latin1]{inputenc}
\usepackage{a4wide}
\usepackage{amsmath}
\usepackage{color}
\usepackage{graphicx}
\usepackage{amssymb}

\newcommand {\beq}{\begin{equation}}
\newcommand {\eeq}{\end{equation}}
\newcommand {\bea}{\begin{eqnarray}}
\newcommand {\eea}{\end{eqnarray}}
\newcommand {\nn}{\nonumber \\}

\newcommand {\K}{{\bf K}}
\newcommand {\I}{{\bf I}}

\newcommand {\m}{\mu}
\newcommand {\n}{\nu}
\newcommand {\pl}{\partial}

\newcommand {\al}{\alpha}
\newcommand {\be}{\beta}

\newcommand {\la}{\lambda}
\newcommand {\La}{\Lambda}

\newcommand {\om}{\omega}

\newcommand {\ep}{\epsilon}

\newcommand {\e} {\mbox{e}}
\newcommand {\na}{\nabla}
\newcommand {\del}  {\delta}

\newcommand {\mn}{{\mu\nu}}

\newcommand {\half}{ {\frac{1}{2}} }

\newcommand {\Lcal}{{\cal L}}

\newcommand {\Dcal}{{\cal D}}
\newcommand {\Ecal}{{\cal E}}

\def\overleftarrow#1{\vbox{\ialign{##\crcr
 $\leftarrow$\crcr\noalign{\kern-1pt\nointerlineskip}
 $\hfil\displaystyle{#1}\hfil$\crcr}}}

\newcommand {\ktil} {{\tilde k}}

\newcommand {\ptil}{{\tilde p}}

\newcommand {\Lhat}{{\hat L}}

\newcommand {\delh} {{\hat \delta}}

\newcommand {\intp} {{\int \frac{d^4p}{(2\pi)^4}}}
\newcommand {\intpE} {{\int \frac{d^4p_E}{(2\pi)^4}}}

\newcommand {\ra} {\rightarrow}

\newcommand {\pr}   {{\quad .}}
\newcommand {\com}  {{\quad ,}}
\newcommand {\q}    {\quad}

\newcommand {\PL}   {Phys.Lett.}
\newcommand {\PR}   {Phys.Rev.}
\newcommand {\PRL}   {Phys.Rev.Lett.}

\newcommand {\PTP}  {Prog.Theor.Phys.}

\newcommand {\Pla} {\frac{{\tilde p}}{\omega}}

\newcommand {\Tev} {\frac{{\tilde p}}{T}}

\begin{document}
%
\title{%
  Renormalization Group Flow and the Dark Energy Problem
}
%
\author{%
  Shoichi Ichinose\footnote{Email address: ichinose@u-shizuoka-ken.ac.jp}
}
%
%
\maketitle
\begin{center}\emph{
Laboratory of Physics, School of Food and Nutritional Sciences, 
University of Shizuoka\\
Yada 52-1, Shizuoka 422-8526, Japan
}\end{center}
\begin{abstract}{
Casimir energy is calculated for 5D scalar theory
in the {\it warped} geometry. 
A new regularization, 
called {\it sphere lattice regularization}, is taken. 
The regularized configuration is {\it closed-string like}. 
We numerically evaluate $\La$(4D UV-cutoff), $\om$(5D bulk curvature, 
extra space UV-boundary parameter)
and $T$(extra space IR-boundary parameter) dependence of Casimir energy. 
5D Casimir energy is {\it finitely} obtained after the {\it proper renormalization
procedure.} 
The {\it warp parameter} $\om$ suffers from the {\it renormalization effect}. 
Regarding Casimir energy as the main contribution to the cosmological term, 
we examine the dark energy problem. 
                   }
\end{abstract}

{\bf 1. Introduction}
In the quest for the unified theory, the higher dimensional (HD) approach 
is a fascinating one from the geometrical point. Historically the initial successful 
one is the Kaluza-Klein model, which unifies the photon, graviton and dilaton
from the 5D space-time approach. 
The HD theories
, however, generally have the serious defect as the quantum field
theory(QFT) : un-renormalizability. 
The HD quantum field theories, at present, 
are not defined within the QFT. 

In 1983, the Casimir energy in the Kaluza-Klein theory was calculated 
by Appelquist and Chodos\cite{P18_AC83}. They took the cut-off ($\La$) regularization and 
found the quintic ($\La^5$) divergence and the finite term. The divergent 
term shows the {\it unrenormalizability} of the 5D theory, but the finite term looks 
meaningful\cite{P18_SI85PLB} 
and, in fact, is widely regarded as the right vacuum energy 
which shows {\it contraction} of the extra axis. 

In the development of the string and D-brane theories, a new approach 
to the renormalization group was found. It is called {\it holographic renormalization}.  
We regard the renormalization group flow as a curve 
in the bulk. The flow goes along the extra axis. 
The curve is derived as a dynamical equation 
such as Hamilton-Jacobi equation. 
It originated from the AdS/CFT correspondence. 
Spiritually the present basic idea overlaps with this approach.

{\bf 2. Casimir Energy of 5D Scalar Theory}
In the warped geometry,   
$
ds^2=\frac{1}{\om^2z^2}(\eta_\mn dx^\m dx^\n+{dz}^2)
$
, we consider the 5D massive {\it scalar} theory with $m^2=-4\om^2~(<0)$. 
$
\Lcal=\sqrt{-G}(-\half \na^A\Phi\na_A\Phi-\half m^2\Phi^2)
$
. The Casimir energy $E_{Cas}$ is given by
\bea
\e^{-T^{-4}E_{Cas}}=\left.\int\Dcal\Phi\exp\{i\int d^5X\Lcal\}\right|_{\mbox{Euclid}}
=\exp\sum_{n,p}\{-\half\ln (p_E^2+M_n^2) \}
,\ p_E^2\equiv p_1^2+p_2^2+p_3^2+p_4^2,
\label{P18_KKexp3}
\eea 
where $M_n$ is the eigenvalues of the following operator. 
\bea 
\{s(z)^{-1}\Lhat_z+{M_n}^2\}\psi_n(z)=0\com\q 
\Lhat_z\equiv\frac{d}{dz}\frac{1}{(\om z)^3}\frac{d}{dz}
                           -\frac{m^2}{(\om z)^5}
\com
\label{P18_KKexp8}
\eea
where $s(z)=\frac{1}{(\om z)^3}$. 
Z$_2$ parity is imposed as:\ 
$
\psi_n(z)=-\psi_n(-z)\ \mbox{for}\ P=-\ ;\ 
\psi_n(z)=\psi_n(-z)\ \mbox{for}\ P=+ 
$. 
The expression (\ref{P18_KKexp3}) is the familiar one 
of the Casimir energy. 
It is re-expressed in a {\it closed} form
using the heat-kernel method and the propagator as follows. 
First we can express it, using the heat equation solution, 
as ($\om/T=\e^{\om l}$, $l$ is the periodicity in $y$-coordinate:\ 
$y\ra y+2l,\om |z|=\e^{\om |y|}$), 
\bea
\e^{-T^{-4} E_{Cas}}
=(\mbox{const})\times\exp \left[ 
T^{-4}\intp 2\int_{0}^{\infty}\half\frac{dt}{t}\mbox{Tr}~H_{p}(z,z';t) 
                          \right] \com\nn
\mbox{Tr}~H_p(z,z';t)=\int_{1/\om}^{1/T}s(z)H_p(z,z;t)dz\com\q
\{\frac{\pl}{\pl t}-(s^{-1}\Lhat_z-p^2) \}H_p(z,z';t)=0
\pr
\label{P18_HKA2}
\eea
The heat kernel $H_p(z,z';t)$ is formally solved, using the
Dirac's bra and ket vectors $(z|, |z)$, as\ \ 
$
H_p(z,z';t)=(z|\e^{-(-s^{-1}\Lhat_z+p^2)t}|z')
$. 
We here introduce the position/momentum propagators $G^{\mp}_p$:\ 
$
G^\mp_p(z,z')\equiv
\int_0^\infty dt~ H_p(z,z';t)
$. 
They satisfy the following differential equations of {\it propagators}.
\bea
(\Lhat_z-p^2s(z))G^{\mp}_p(z,z')=
\left\{
\begin{array}{ll}
\ep(z)\ep(z')\delh (|z|-|z'|) & \mbox{\ P=}-1 \\
\delh (|z|-|z'|) & \mbox{\ P=}1 
\end{array}
        \right.
        \com\q p^2=p_\mu p^\mu=-p_0^2+p_1^2+p_2^2+p_3^2
\label{P18_HKA7}
\eea

$G_p^\mp$ can be expressed in a {\it closed} form.
Taking the {\it Dirichlet} condition at all fixed points, the expression
for the fundamental region ($1/\om \leq z\leq z'\leq 1/T$) is given by
\bea
G_p^\mp(z,z')=\mp\frac{\om^3}{2}z^2{z'}^2
\frac{\{\I_0(\Pla)\K_0(\ptil z)\mp\K_0(\Pla)\I_0(\ptil z)\}  
      \{\I_0(\Tev)\K_0(\ptil z')\mp\K_0(\Tev)\I_0(\ptil z')\}
     }{\I_0(\Tev)\K_0(\Pla)-\K_0(\Tev)\I_0(\Pla)},
\label{P18_HKA12}
\eea 
where $\ptil\equiv\sqrt{p^2},p^2\geq 0$. $I_0$ and $K_0$ are modified Bessel functions. 
We can express Casimir energy as, 
\bea
-E^{\La,\mp}_{Cas}(\om,T)
=\left.\intpE\right|_{\ptil\leq\La}\int_{1/\om}^{1/T}dz~F^\mp(\ptil,z), 
F^\mp(\ptil,z)
=\frac{2}{(\om z)^3}\int_\ptil^\La\ktil~ G^\mp_k(z,z)d\ktil
,
\label{P18_HKA13}
\eea
where $\ptil=\sqrt{p_E^2}$. The momentum symbol $p_E$ indicates Euclideanization. 
Here we introduce the UV cut-off parameter $\La$ in the 4D momentum space.

{\bf 3. UV and IR Regularization and Evaluation of Casimir Energy}
The integral region of the above equation (\ref{P18_HKA13}) is displayed, 
in Fig.1, as a rectangle. 
In the figure, we introduce the regularization cut-offs for the 4D-momentum integral, 
$\m\leq\ptil\leq\La$. 
For simplicity, we take
the following IR cutoff of 4D momentum\ :\ 
$\m=\La\cdot\frac{T}{\om}=\La \e^{-\om l}$ . 
\begin{figure}
\caption{
Rectangle region of (z,$\ptil$) for the integration (\ref{P18_HKA13}). 
The hyperbolic curve 
was proposed\cite{P18_RS01}. 
}
\includegraphics[height=8cm]{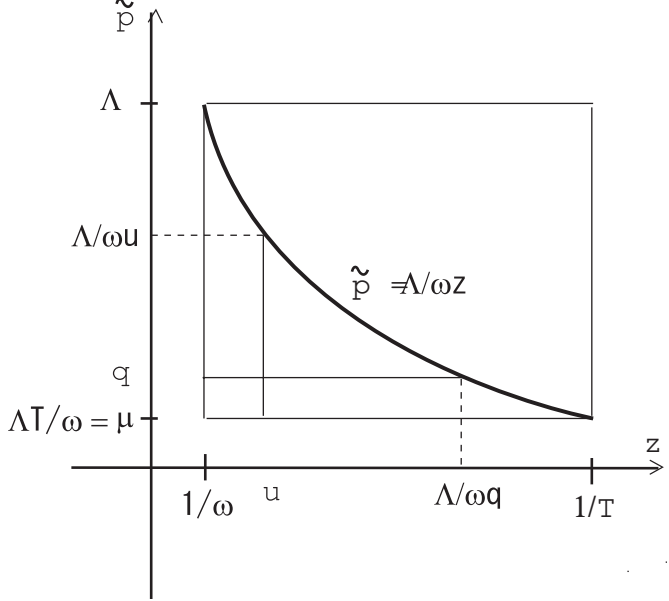}
\label{zpINTregionW}
\end{figure}
\begin{figure}
\caption{
Region of ($\ptil$,z) for the integration (present proposal). 
}
\includegraphics[height=8cm]{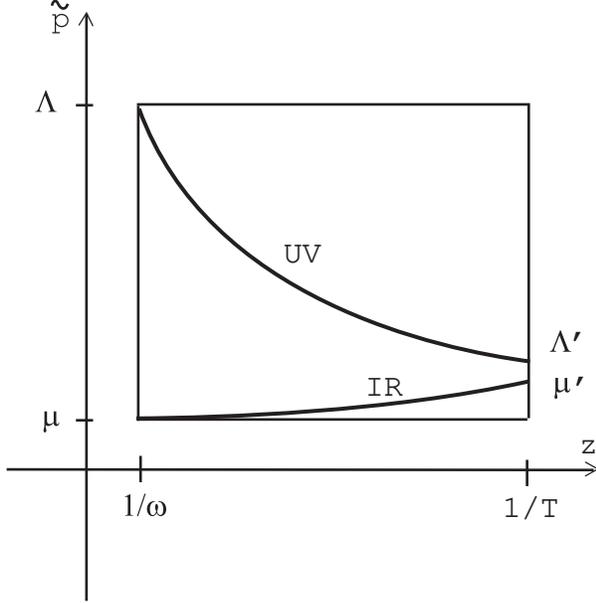}
\label{zpINTregionW2}
\end{figure}
Importantly, (\ref{P18_HKA13}) shows the {\it scaling} behavior for large values of $\La$ and $1/T$. 
From a {\it close} numerical analysis, 
we have confirmed\ :\ (4A)\ 
$E^{\La,-}_{Cas}(\om,T)=\frac{2\pi^2}{(2\pi)^4}\times\left[ -0.0250 \frac{\La^5}{T} \right]$. 
The $\La^5$-divergence, (4A), shows the notorious problem
of the higher dimensional theories. 
We have proposed 
an approach to solve this problem 
and given a legitimate explanation within the 5D QFT\cite{P18_IM0703,P18_SI0801}. 
See Fig.2. The IR and 
UV cutoffs {\it change} along the etra axis. 
Their $S^3$-radii are given by $r_{IR}(z)=1/\ptil_{IR}(z)$ and 
$r_{UV}(z)=1/\ptil_{UV}(z)$. 
The 5D volume region bounded by $B_{UV}$ and $B_{IR}$ is the integral region 
 of the Casimir energy $E_{Cas}$. (
We call this regularization {\it sphere lattice regularization}
because one big 4D-ball (radius $r_{IR}$) are made of 
many small 4D-balls (cells, radius $r_{UV}$). See Fig.12 of Ref.\cite{P18_SI0812}
                                    ) 
The forms of $r_{UV}(z)$ and $r_{IR}(z)$ can be
determined by the {\it minimal area principle}:\ 
$
3+\frac{4}{z}r'r-\frac{r''r}{{r'}^2+1}=0
, r'\equiv\frac{dr}{dz}
, r''\equiv\frac{d^2r}{dz^2}
, 1/\om\leq z\leq 1/T
$. 
We have confirmed, by {\it numerically} solving the above differential eqation 
(Runge-Kutta), those curves 
that show the flow of renormalization really appear. 
The results imply the {\it boundary conditions} 
determine the property of the renormalization flow.

{\bf 4. Weight Function and the Meaning}
We consider another approach which respects 
the {\it minimal area principle}. 
Let us introduce, instead of restricting the integral region, 
a {\it weight function} $W(\ptil,z)$ in the ($\ptil,z$)-space  
for the purpose of {\it suppressing} UV and IR divergences of the Casimir Energy. 
\bea
-E^{\mp~W}_{Cas}(\om,T)\equiv\intpE\int_{1/\om}^{1/T}dz~ W(\ptil,z)F^\mp (\ptil,z)\com\q
\ptil=\sqrt{p_4^2+p_1^2+p_2^2+p_3^2}\com\nn
\left\{
\begin{array}{cc}
(N_1)^{-1}\e^{-(1/2) \ptil^2/\om^2-(1/2) z^2 T^2}\equiv W_1(\ptil,z),\ N_1=1.711/8\pi^2 & \mbox{elliptic suppr.}\\
(N_{2})^{-1}\e^{-\ptil zT/\om}\equiv W_2(\ptil,z),\ N_2=2\frac{\om^3}{T^3}/8\pi^2                   & \mbox{hyperbolic suppr.1}\\
(N_{8})^{-1}\e^{-1/2 (\ptil^2/\om^2+1/z^2T^2)}\equiv W_8(\ptil,z),\ N_8=0.4177/8\pi^2 & \mbox{reciprocal suppr.1}\\
\end{array}
           \right.
\label{P18_uncert1}
\eea
where 
$F^\mp(\ptil,z)$ are defined in (\ref{P18_HKA13}). 
They (except $W_2$) give, after normalizing the factor $\La/T$, {\it only} the {\it log-divergence}. 
\bea
E^W_{Cas}/\La T^{-1} =-\al \om^4\left( 1-4c\ln (\La/\om) -4c'\ln (\La/T) \right) 
\com
\label{P18_uncert1c}
\eea
where the numerical values of $\al, c$ and $c'$ are obtained depending on the choice of 
the weight function\cite{P18_SI0812}. 
This means the 5D Casimir energy is {\it finitely} obtained by the ordinary 
renormalization of the warp factor $\om$. (See the final section.)


In the previous work\cite{P18_SI0801}, we have presented the following idea to 
define the weight function $W(\ptil,z)$. In the evaluation (\ref{P18_uncert1}), 
the $(\ptil,z)$-integral is over the rectangle region shown in Fig.1 
(with $\La\ra\infty$ and $\m\ra 0$). 
Following Feynman\cite{P18_Fey72}, 
we can replace the integral by the {\it summation} over {\it all possible pathes} $\ptil(z)$. 
\bea
-E^{W}_{Cas}(\om,T)=\int\Dcal\ptil(z)\int_{1/\om}^{1/T}dz~S[\ptil(z),z],
S[\ptil(z),z]=\frac{2\pi^2}{(2\pi)^4}\ptil(z)^3 W(\ptil(z),z)F^\mp (\ptil(z),z).
\label{P18_weight1a}
\eea
There exists the {\it dominant path} $\ptil_W(z)$ which is 
determined by the minimal principle
: 
$\del S=0$.\ 
Dominant Path\ 
$\ptil_W(z)\ :\ \q$ 
$\frac{d\ptil}{dz}=$ 
$
-\frac{\pl\ln(WF)}{\pl z}
$/
$(                       
\frac{3}{\ptil}+\frac{\pl\ln (WF)}{\pl\ptil}
  )$.  
Hence it is fixed by the weight function $W(\ptil,z)$.  
On the other hand, there exists another independent path: the minimal surface  
curve $r_g(z)$.\  
Minimal Surface Curve\ $r_g(z)$\ :\ 
$3+\frac{4}{z}r'r-\frac{r''r}{{r'}^2+1}=0$,\ 
$\frac{1}{\om}\leq z\leq \frac{1}{T}$. 
It is obtained by the {\it minimal area principle}:\ $\del A=0$ where 
\bea
ds^2=(\del_{ab}+\frac{x^ax^b}{(rr')^2} )
\frac{dx^a dx^b}{\om^2 z^2}\equiv g_{ab}(x)dx^adx^b ,
A=\int\sqrt{g}~d^4x
=\int_{1/\om}^{1/T}\frac{  
                         \sqrt{{r'}^2+1}~r^3} 
                        {\om^4z^4}
dz.
\label{P18_weight3}
\eea 
Hence $r_g(z)$ is fixed by the {\it induced geometry} $g_{ab}(x)$. 
Here we put the {\it requirement}\cite{P18_SI0801}:\ (4A)\ 
$\ptil_W(z)=\ptil_g(z)$, 
where $\ptil_g\equiv 1/r_g$. This means the following things. 
We {\it require} 
the dominant path coincides with the minimal surface line $\ptil_g(z)=1/r_g(z)$ which is 
defined independently of $W(\ptil,z)$. 
$W(\ptil,z)$ is defined here by 
the induced geometry $g_{ab}(x)$. 
In this way, we can connect the integral-measure over the 5D-space with the geometry. 
We have confirmed the coincidence by the 
numerical method. 

In order to most naturally accomplish 
the above requirement, we can go to a 
{\it new step}. Namely, 
we {\it propose} to {\it replace} the 5D space integral with the weight $W$, (\ref{P18_uncert1}), 
by the following {\it path-integral}. We 
{\it newly define} the Casimir energy in the higher-dimensional theory as follows.  
\bea
-\Ecal_{Cas}(\om,T,\La)=
\int_{1/\La}^{1/\m}d\rho\int
{
\begin{array}{l}
r(\om^{-1})\\
=r(T^{-1})\\
=\rho
\end{array}
}
\prod_{a,z}\Dcal x^a(z)F(\frac{1}{r},z)
\exp\left[ 
-\int_{1/\om}^{1/T}\frac{\sqrt{{r'}^2+1}~r^3}{2\al'\om^4z^4}dz
    \right],
\label{P18_weight5}
\eea 
where $\m=\La T/\om$ and the limit $\La T^{-1}\ra \infty$ is taken. 
The string (surface) tension parameter $1/2\al'$ is introduced.
 (Note: Dimension of $\al'$ is [Length]$^4$. ) 
$F(\ptil,z)$ is defined in (\ref{P18_HKA13}) and 
represents the contribution from 
the {\it field-quantization} of the bulk scalar (or EM) fields. 
(This proposal is shown to be valid in arXiv:1004.2573, 1010.5558.)

{\bf 5. Discussion and Conclusion}
When $c$ and $c'$ in (\ref{P18_uncert1c}) are sufficiently small 
we find the renormalization group function for the warp factor $\om$ as  
\bea
\om_r=\om (1-c\ln (\La/\om)-c'\ln (\La/T) )\ ,\ 
\be\equiv \frac{\pl}{\pl(\ln \La)}\ln \frac{\om_r}{\om}=-c-c'
\pr
\label{P18_conc2}
\eea
No local counterterms are necessary. 

Through the Casimir energy calculation, in the higher dimension, we find a way to 
quantize the higher dimensional theories within the QFT framework. 
The quantization {\it with respect to the fields} (except the gravitational fields $G_{AB}(X)$) 
is done in the standard way. After this step, the expression has the summation 
{\it over the 5D space(-time) coordinates or momenta} 
$\int dz\prod_adp^a$. We have proposed that this summation should be replaced by 
the {\it path-integral} $\int \prod_{a,z}\Dcal p^a(z)$ with the {\it area} action (Hamiltonian) 
$A=\int\sqrt{\det g_{ab}}~d^4x$ where $g_{ab}$ is the {\it induced} metric on the 4D surface. 
This procedure says the 4D momenta 
$p^a$ (or coordinates $x^a$) are {\it quantum statistical} operators and 
the extra-coordinate $z$ is the inverse temperature (Euclidean time). 
We recall the similar situation occurs in the standard string approach. 
The space-time coordinates obey some uncertainty principle\cite{P18_Yoneya87}.  

Recently the dark energy (as well as the dark matter) in the universe is a hot subject. 
It is well-known that the dominant candidate is the cosmological term. 
The cosmological
constant $\la$ appears as:\ (5A)\ 
$R_\mn-\half g_\mn R-\la g_\mn =T_\mn^{matter},
S=\int d^4x \sqrt{-g}\{\frac{1}{G_N}(R+\la) \}
+\int d^4x \sqrt{-g}\{\Lcal_{matter}\}, 
g=\mbox{det}~g_{\mn}$ 
. 
We consider here the 3+1 dim Lorentzian space-time ($\mu,\nu=0,1,2,3$). 
The constant $\la$ observationally takes the value\ :\ (5B)\ 
$\frac{1}{G_N}\la_{obs}\sim \frac{1}{G_N{R_{cos}}^2}\sim m_\n^4\sim (10^{-3} eV)^4
,
\la_{obs}\sim \frac{1}{R_{cos}^{~2}}\sim 4\times 10^{-66}(eV)^2$,  
where $R_{cos}\sim 5\times 10^{32}\mbox{eV}^{-1}$ is the cosmological size (Hubble length), $m_\n$ is the neutrino mass.
On the other hand, we have theoretically so far\ :\ (5C)\ 
$\frac{1}{G_N}\la_{th}\sim \frac{1}{{G_N}^2}={M_{pl}}^4\sim (10^{28} eV)^4$. 
We have the famous huge discrepancy factor\ :\ (5D)\ 
$\frac{\la_{th}}{\la_{obs}}\sim N_{DL}^{~2}, N_{DL}\equiv M_{pl}R_{cos}\sim 6\times 10^{60}$, 
where $N_{DL}$ is here introduced as the Dirac's large number. 
If we use the present result, we can obtain a natural choice of 
$T, \om$ and $\La$
as follows. By identifying 
$T^{-4}E_{Cas}=-\al_1\La T^{-1}\om^4/T^4$ with 
$\int d^4x\sqrt{-g}(1/G_N)\la_{ob}=R_{cos}^{~2}(1/G_N)$, we 
obtain the following relation:\ (5E)\  
$N_{DL}^{~2}=R_{cos}^{~2}\frac{1}{G_N}=-\al_1 \frac{\om^4\La}{T^5}$. 
The warped (AdS$_5$) model predicts the cosmological constant {\it negative}, 
hence we have interest only in its absolute value.
We take the following choice for $\La$ and $\om$\ :\ (5F)\ 
$\La=M_{pl}\sim 10^{19}GeV, 
\om\sim
 \frac{1}{\sqrt[4]{G_N{R_{cos}}^2}}=\sqrt{\frac{M_{pl}}{R_{cos}}}\sim m_\n\sim 10^{-3}\mbox{eV}$. 

As shown above, we have the standpoint that the cosmological constant is mainly made from 
the Casimir energy.  
We do not yet succeed in obtaining the value $\al_1$ negatively, but
succeed in obtaining  
the finiteness of the cosmological constant and its gross absolute value. 
The smallness of the value is naturally explained by the renormalization group flow. 
Because we already know the warp parameter $\om$ {\it flows} (\ref{P18_conc2}), 
the $\la_{obs}\sim 1/R_{cos}^2\propto \om^4$, says that the {\it smallness of the cosmological constant comes from 
the renormalization group flow} for the non asymptotic-free case ($c+c'<0$ in (\ref{P18_conc2})). 

The IR parameter $T$, the normalization factor $\La/T$ in (\ref{P18_uncert1c}) and the IR cutoff 
$\mu=\La\frac{T}{\om}$ are given by\ :\ (5G)\  
$T=R_{cos}^{~-1}(N_{DL})^{1/5}\sim 10^{-20}eV, 
\frac{\La}{T}=(N_{DL})^{4/5}\sim 10^{50}, 
\mu=M_{pl}N_{DL}^{-3/10}\sim 1GeV\sim m_N$, 
where $m_N$ is the nucleon mass. 
The degree of freedom of the universe (space-time) 
is given by\ :\ (5H)\  
$\frac{\La^4}{\m^4}=\frac{\om^4}{T^4}=N_{DL}^{~6/5}\sim 10^{74}\sim (\frac{M_{pl}}{m_N})^4$. 

\end{document}